# Optimizing floating guard ring designs for FASPAX N-in-P silicon sensors

Kyung-Wook Shin, Robert Bradford, *Member, IEEE,* Ronald Lipton, Gregory Deptuch, *Senior Member, IEEE,* Farah Fahim, Tim Madden, *Member, IEEE,* and Tom Zimmerman

*Abstract*—FASPAX (Fermi-Argonne Semiconducting Pixel Array X-ray detector) is being developed as a fast integrating area detector with wide dynamic range for time resolved applications at the upgraded Advanced Photon Source (APS.) A burst mode detector with intended 13 $MHz$ image rate, FASPAX will also incorporate a novel integration circuit to achieve wide dynamic range, from single photon sensitivity to $10^5$ x-rays/pixel/pulse. To achieve these ambitious goals, a novel silicon sensor design is required. This paper will detail early design of the FASPAX sensor. Results from TCAD optimization studies, and characterization of prototype sensors will be presented.

## I. Introduction

SEMICONDUCTOR X-ray detectors with a direct-conversion scheme, where the X-ray photons are converted to electrical signals within the semiconductor, were first used for synchrotron radiation in the early 1980s [1–3]. Such detectors were proceeded by much more delicate photon counting detectors since late 1990s [4, 5]. However, photon counting detectors suffer from pulse-pileup effects due to the necessity of the pulse shaping operation, limiting their count rate to the *Mcps/pixel* scale at most, in 2D pixellated detectors [6]. These count rate limitations constrain many techniques common synchrotron techniques, such as coherent diffractive imaging, X-ray phase contrast imaging, X-ray photon correlation spectroscopy, and scanning probe imaging, etc. [7]

### A. APS Upgrade

The APS (Advanced Photon Source) is preparing for a major facility upgrade that will include installation of a new diffraction limited storage ring. Tab. I summarizes key parameters of the APS upgrade.

In short, the APS upgrade will provide a brighter beams with 2 times higher repetition rate and smaller horizontal emittance. As many photon counting X-ray detectors require beamline attenuation to avoid pulse pileup even

Manuscript received January 4, 2016. This work was supported in part by the U.S. Department of Energy, Office of Science, Office of Basic Energy Sciences, under Contract No. DE-AC02-06CH1135.

Kyung-Wook Shin, Robert Bradford, and Tim Madden are with the Argonne National Laboratory, X-ray Science Division, Argonne, IL 60439 USA (telephone: 630-252-1256, e-mail: kshin@aps.anl.gov).

Ronald Lipton, Gregory Deptuch, Farah Fahim, and Tom Zimmerman are with the Fermi National Laboratory, Particle Physics Division, Batavia, IL 60510-5011 USA (telephone: 630-840-4569, e-mail: deptuch@fnal.gov).

TABLE I: Summary of APS Upgrade Beamline Specifications. Adopted from [7].

| Parameters | Now | After Upgrade |
|---|---|---|
| Beam Energy ($GeV$) | 7 | 6 |
| Beam Current ($mA$) | 100 | 200 |
| Number of Bunches | 24 | 48 |
| Bunch Duration ($ps$) | 34 | 70 |
| Bunch Spacing ($ns$) | 153 | 77 |
| Bunch Rep. rate ($MHz$) | 6.5 | 13 |
| Horizontal Emittance ($pm$) | 3100 | 42 |
| Horizontal Beam Size ($\mu m$) | 265 | 7.4 |
| Horizontal Beam Divergence ($\mu rad$) | 11 | 57 |
| Vertical Emittance ($pm$) | 40 | 42 |
| Vertical Beam Size ($\mu m$) | 10 | 10.9 |
| Vertical Beam Divergence ($\mu rad$) | 3.5 | 3.8 |

before the upgrade, new x-ray area detectors capable of handling the increased count rates of the upgraded beamlines are mandatory. To access the full temporal resolution of the new storage ring, the new detector must achieve a frame rates of 13 $MHz$, consistent with the increased bunch rate. To overcome the limtations of traditional photon counting detectors, we have opted to use a high speed integrating readout circuit for the FASPAX project.

### B. FASPAX Detector

In order to fully use the incrased brightness of the upgraded beamlines, we have taken the integration approach rather than counting impinging photons one by one in FASPAX. Such method was already implemented in active matrix flat panel imagers (AMFPIs) which are widespread in medical X-ray imaging [8] but lacking fast readout method due to limitation of large area electronics performance (field effect mobility of an amorphous silicon thin-film transistor can be 1.45 $cm^2/V \cdot s$ at most [9]) and active matrix readout scheme (reading out line by line.) However, if we read out each pixel simultaneously with a fast readout circuit, it is possible to overcome pulse pileup [10].

On the other hand, such high influx of photons cause a detrimental effect called plasma (delay) effect which is caused by high photo-generated carrier density. Such high density carriers form a "shield" from the sensors bias, thus extracting the photo-generated carriers slower than expected [11–13]. This effect can be averted by simply applying higher bias (i.e. over 500 $V$ on 280-$\mu m$-thick silicon diode) on the sensor [14, 15].

However, applying such high bias leaves the photo-converter in peril due to trap assisted breakdown at the detector silicon wafer termination as well as trap states generated by prolonged X-ray exposure. Thus, proper guard ring implementation on silicon diode X-ray detectors is mandatory. These guard ring structures have been commonly implemented in power electronics since the 1960s [16] for power switching devices and have been implemented in silicon high energy particle detectors and silicon carbide high voltage diode detectors [17–19].

### C. FASPAX Sensor

We are prototyping N-in-P silicon diode detectors (See Figure 2) for use with a high-speed integrating readout. The N-in-P detectors are basically a silicon diode with pixellated n-type contacts in a highly resistive p-type substrate while the back of the wafer has a large p-type implant to provide an ohmic contact to the detector as depicted in Fig. 1.

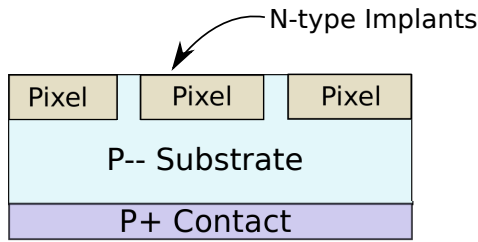

Fig. 1: Basic structure of silicon diode N-in-P detectors.

N-in-P detectors are advantageous over conventional n-type substrate detectors (i.e. P-in-N devices) since n-type implant on almost intrinsic p-type substrate enables collecting electrons which have a bulk mobility (1400 $cm^2/V \cdot s$) twice that of holes (450 $cm^2/V \cdot s$) at room temperature (300 $^oK$.) Moreover, p-type substrate is known for superior radiation hardness over any n-type substrate devices and the fabrication does not require two side lithography which is mandatory for N-in-N device fabrication [20–22].

Current prototypes, fabricated from Novati Technologies™, showed an unexpectedly low break-down bias of -100 to -120 $V$ which we have investigated with Silvaco™ ATLAS TCAD (Device 3D) [23] to resolve the design problem. This paper also reports the TCAD investigation of a revised guard ring design which has been included in the next Novati mask tape-out which was submitted in the end of September 2015.

## II. Guard Ring Breakdown

### A. Sample Preparation and Measurement

FASPAX N-in-P prototypes were fabricated on high resistivity float zone silicon wafers (Fig. 2) with equivalent doping concentration of $10^{12}$ $cm^{-3}$ boron acceptors. Each square shaped pixel has dimension of 100 $\mu m$ pitch (as tested) and the entire 32 by 32 pixel matrix is enclosed by a 100-$\mu m$-wide current collection ring (CCR) separated by 15 $\mu m$ and a 5 $\mu m$ p-stop implant between pixels and the CCR. We picked 15 guard ring prototypes for guard ring breakdown test which have gradually increasing guard ring width and space between guard rings. The guard ring width starts from 15 $\mu m$ (innermost guard ring) and increase by 1.5 $\mu m$ by the guard ring number, i.e. $13.5 + 1.5 \times N$ where $N$ is an integer between 1 and 15. The p-stops between guard rings were implemented with a constant width of 10 $\mu m$ which was located exactly at the middle of the spacings between guard rings. The detector design was heavily influenced from AGIPD for European XFEL [24].

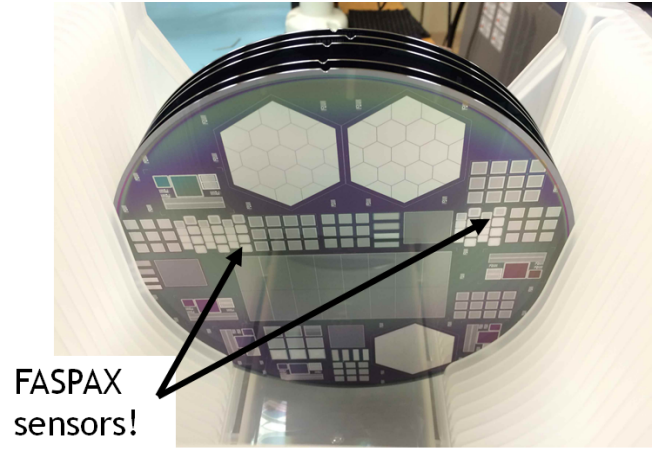

Fig. 2: Wafer snapshot of N-in-P configuration FASPAX detectors.

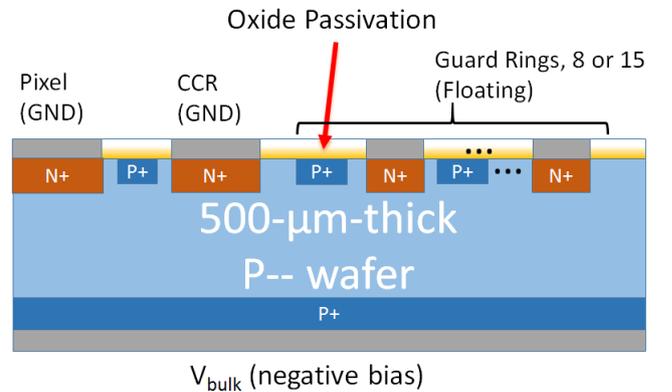

Fig. 3: I-V test scheme on guard ring breakdown.

Each guard ring electrode was 25-$\mu m$-wide with aliminum overhangs on the 0.5-$\mu m$-thick passvation oxide layer. The overhang towards the pixel side grows with $1.0 + 1.0 \times N$ relation while the other side stays at 5 $\mu m$. Of course, the space between overhanging electrodes was also filled by 0.5-$\mu m$-thick oxide.

The breakdown test was performed with a Keithley 237 High Voltage Source-Measure Unit, connected to a

PC via GPIB interface to record the current response with a customized readout software, based on National Instruments™ LabView™. The entire measurement set up and data readout software were provided by the Silicon Detector (SiDet) Facility at Fermi National Accelerator Laboratory (FNAL.)

As depicted in Fig. 3, the N-in-P detector was biased with negative bias at the bulk electrode ($V_{bulk}$) while its pixel and CCR were both connected to ground bias to mimic detector operation when it is integrated with the ASIC reaout circuitry. The bulk electrode bias sweep was programmed from 0 $V$ to -1100 $V$ with -10 $V$ of increment to induce breakdown.

### B. Measurement - Early Breakdown

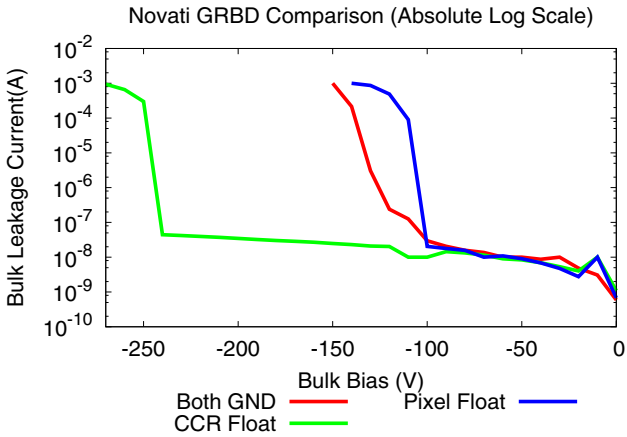

Fig. 4: Low bias breakdown from the first prototype N-in-P FASPAX detector.

The guard ring breakdown (GRBD) was found much earlier (-100 to -120 $V$) than expected as seen in Fig. 4 when the pixel and the CCR electrodes were both grounded. Leaving the pixel electrode floating resulted similarly resulted in low breakdown bias of -100 $V$ while floating CCR with grounded pixel pushes the breakdown down to around -240 $V$ which is still insufficient to avoid the plasma delay effect.

Since we can manipulate GRBD characteristics by floating the CCR, it is easy to assume that the first guard ring p-contact plays the most critical role in GRBD. Thus, we implemented a numerical analysis approach with TCAD tools to support the assumption and to provide insight on guard ring design for the next batch of prototypes. Following sections depict TCAD simulation results on the current design and the next generation prototypes.

### III. TCAD Simulation Results

#### A. Preparing TCAD Model

The TCAD simulation was preformed with Silvaco™ Atlas 5.20.2.R package (Device3D) on an HP™ Z820 workstation equipped with 3.2 $GHz$, 96 logical CPUs with Intel® Hyper-Threading Technology. The guard ring simulation model was taken from half the last pixel to the end of the wafer termination (1260 $\mu m$ from the center of last pixel to wafer termination,) similar to Fig. 3 with impurity refined mesh generation, sensitivity of $10^{2.5}$ and transition value of 10.0, provided by Silvaco™ Devedit3D 2.8.21.R [25]. The simulation models were shrunk down to 1 $\mu m$ of thickness (Z-axis) to reduce simulation time which usually takes 48 to 60 hours to finish. Note that Atlas 2D simulation modules assume the device thickness (Z-axis) of 1 $\mu m$, thus such shrunken device model does not compromise simulation accuracy [26].

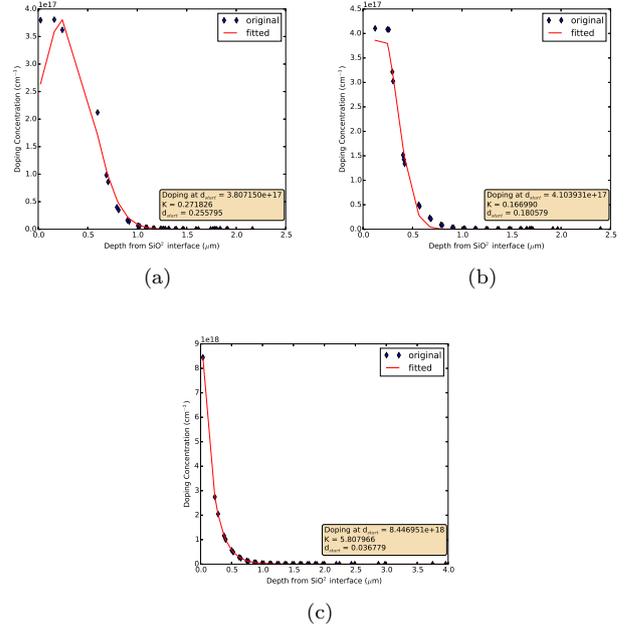

Fig. 5: Least square minimization fit results on p-stop SIMS profiles: vertical axis fit for p-stop implant concentration (a), lateral axis fit for p-stop implant concentration (b), and p-implant at bulk electrode (c).

Silicon-silicon dioxide (passivation oxide) interface trap density was assigned a rather typical value of $8.8 \times 10^{11}$ $cm^{-2}$ while neglecting the trap states at the wafer termination surface. We enabled the concentration dependent Schottkly-Read Hall model (Scharfetter relation [27], CONSRH) the parallel electric field dependence model (FLDMOB,) the auger recombination model (AUGER,) the band gap narrowing model (BGN,) Lombardi's mobility model (CVT,) the Kane band-to-band model (BTBT,) and Fermi-Dirac statistics (FERMI) with default parameters [26].

Meanwhile, the vertical portion of the impurity concentration was curve-fitted to secondary ion mass spectrometry (SIMS) analysis was performed by Novati Technologies Inc. via SiDet at FNAL. The fitting (Fig. 5 (a) and (c)) was performed through a least square minimization method on normalized SIMS data using Python open-source general purpose script language accompanied with NumPy open-source numerical library module. The residual function for the vertical p-stop implant profile fit was a Gaussian, depicted by Equation 1 while the bulk contact implant was curve fitted with an exponential residual from Eq. 2 [25]. The lateral axis profile fitting was performed on Silvaco™ Athena 2D process simulation results, provided by SiDet at FNAL, with Gaussian residual as depicted in Fig. 5 (b). Note that the fitting parameters in the residual functions, Eq. 1 and 2, are denoted as $K$. Likewise, the doping profiles for n-type (phosphorus) implants have been curve fitted over SIMS data and Athena simulator results as depicted in Fig. 6. Tab. II summarizes the least square minimization results.

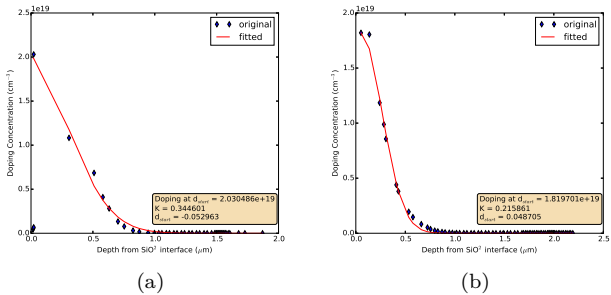

Fig. 6: Least square minimization fit on phosphorus impurity. (a) vertical direction, (b) lateral direction.

$$\text{error} = \text{data} - \exp\left(-\frac{d^2}{2K^2}\right) \quad (1)$$

$$\text{error} = \text{data} - \exp\left(-dK\right) \quad (2)$$

To implement high-purity bulk silicon substrate, minority carrier lifetime parameters (TAUN0 and TAUP0) were adjusted for both electron and holes as 1 $ms$ [28, 29]. Also, reference doner and acceptor concentrations for the Scharfetter relation (NSRHN and NSRHP) were also adjusted to $10^{16}$ $cm^{-3}$. However, the substrate base impurity concentration was left at zero, a completely intrinsic semiconductor, to avoid an Atlas numerical solver convergence issue. Lastly, Selberherr's impact ionization model (SELB) with default parameters was implemented for avalanche breakdown simulation.

The Atlas numerical solver was configured to adopt the Newton-Richardson method (AUTONR) to accelerate processing and increased trapping algorithm steps to 200 while the Newton method iteration limit was increased to 50 from the default parameter of 15 to avoid convergence problems. Also, the carrier concentration convergence parameter (CLIMIT) was fixed to $10^{-4}$ since breakdown is expected within specified the bias condition [26]. Although the N-in-P detectors are known to be collect electrons, we enabled hole solvers to improve accuracy at breakdown.

It is intrinsically impossible to expect any solution from Atlas when a floating electrode is making contact with any semiconductor region. Thus, we implemented $20 \times 10^{20}$ $\Omega$ lumped element resistances to guard ring electrodes which ensures the leakage current through such guard rings as low as $10^{-18}$ $A$ range at most, during bulk bias voltage sweep in simulation.

*B. Spotting the Breakdown*

The simulation results were obtained with TonyPlot 3.8.52.R. As the breakdown sweep in Fig. 7 (a) indicates, the guard ring breakdown can be observed at $V_{\text{bulk}}$ = -600 $V$ which is marginally higher than the experimental data in Fig. 4. This discrepancy may have stemmed from the lack of substrate impurity and under estimated oxide interface trap density. Also, the lateral profile of p-stop impurity was not exactly curve fitted onto experimental data but was rather based on process simulation. The "Function 1" in Fig. 7 (a) is absolute data set of leakage current through the bulk electrode, i.e. $|I_{bulk}|$.

On the other hand, the breakdown was taking place at the first p-stop implant junction, which is effectively reverse biased N-i-P diode, as depicted in Fig. 7 (b). The lateral current density in the Fig. 7 (b) was extracted from a cut-line which is 1 $nm$ beneath the silicon dioxide passivation when the $V_{\text{bulk}}$ bias was reached at -900 $V$.

It can be noted that the reverse bias between CCR n-type contact implant and the first guard ring p-stop implant was 60 $V$ as depicted in the surface potential distribution in Fig. 8 (a) which was also taken from 1 $nm$ below the silicon dioxide passivation. In addition, the potential drop decreases as the distance between the CCR and first guard ring increases. On the bright side, the 15 guard ring design was capable of deceasing the top surface potential at the wafer termination to match the $V_{\text{bulk}}$ bias of -600 $V$.

Indeed, the intensity of electric field (as seen in Fig. 8 (b), also taken from 1 $nm$ beneath the silicon dioxide passivation) decreases as the distance from CCR increases. Also, the p-stop implant between the pixel and the CCR is only one third of CCR-to-1st guard ring vicinity. The electric field intensity at the breakdown junction is 0.29 $MV/cm$.

IV. REVISED DESIGN

*A. P-Stop Implant*

Obviously, the CCR to p-stop guard ring vicinity has a N-i-P diode breaking down at 60 $V$ of local reverse bias. As depicted in Fig. 9 (a), the distances between the n-type implant and the p-type implants are 4 $\mu m$. Such breakdown thresholds can be improved by either reducing the implant concentration or expanding the intrinsic layer thickness to reduce the electric field at a given reverse bias. Since controlling the doping concentration of p-stop implants or n-type metal contact implants is limited by the N-in-P detector design and the fabrication procedure (the guard ring p-stops are implanted with pixel p-stops,) we can only play with the distance between the implants.

TABLE II: Summary of SIMS data least squre fitting results for Devedit model generation.

| Parameters | P-stop vertical | P-stop lateral | Bulk electrode | N-implant vertical | N-implant lateral |
|---|---|---|---|---|---|
| Peak Doping Concentration ($cm^{-3}$) | $3.807150 \times 10^{17}$ | $4.103931 \times 10^{17}$ | $8.446951 \times 10^{18}$ | $2.030486 \times 10^{19}$ | $1.819701 \times 10^{19}$ |
| Fitting Model | Gaussian | Gaussian | Exponential | Gaussian | Gaussian |
| Fitting Parameter ($K$, $\mu$m) | 0.271826 | 0.166990 | 5.807966 | 0.344601 | 0.215861 |
| Residual Function Shift ($d_{start}$, $\mu$m) | 0.255795 | 0.180579 | 0.36779 | $-0.052963$ | 0.048705 |

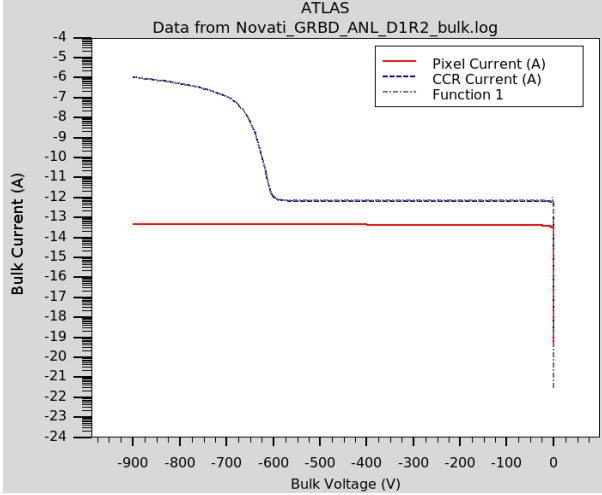

(a)

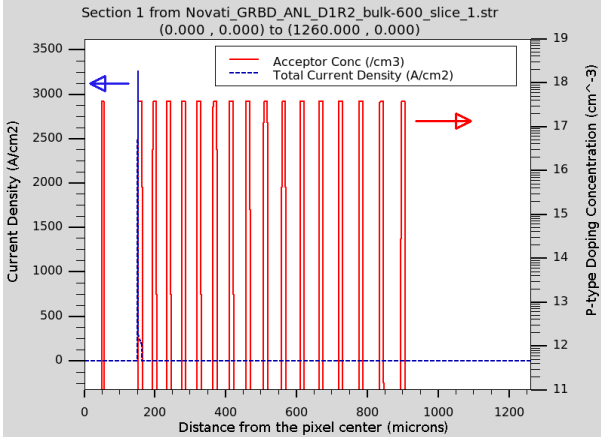

(b)

Fig. 7: (a) Simulated breakdown was spotted at $V_{bulk}$ = -600 $V$ and (b) the first guard ring contact (p-stop contact) is solely contributing the breakdown. Note that the "Function 1" means an absolute value of the bulk current: $|I_{bulk}|$ since $I_{bulk}$ is negative.

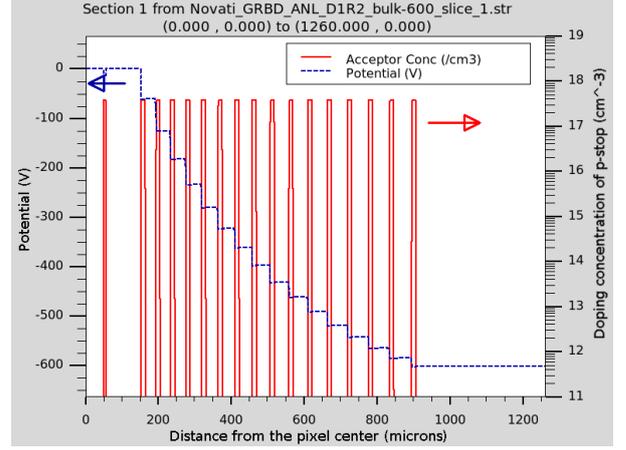

(a)

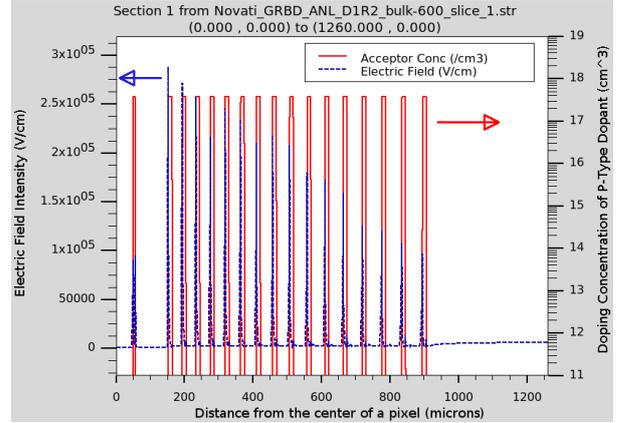

(b)

Fig. 8: (a) Surface potential distribution and (b) electric field at the guard ring junctions.

Thus, we decided to shorten the guard ring p-stop width to 6 $\mu m$ to improve reverse bias strength of N-i-P (n-type, intrinsic, p-type) surface as depicted in Fig. 9 (b) by adding up 0.5 $\mu m$ of extra intrinsic silicon between impurities. Reducing the p-stop width also allowed less space consumption by guard rings. The spacing between guard rings was adjusted to $10.5 + 1.5 \times N$ (where, $N$ is an integer between 1 and 15,) eventually saving 20 $\mu m$ of space from each side. The guard ring electrode width (electrodes sitting on n-type implants) has not been changed from 25 $\mu m$ but the metal width can also be shrunken since the metal electrodes only provide contact pads for external voltage sources and probes. The entire device dimension was also preserved as 1260 $\mu m$.

*B. TCAD Simulation Results - Revised Design*

The revised design shows significant improvement over the previous design by more than doubling the breakdown bulk bias ($V_{bulk}$) of -1350 $V$ as depicted in Fig. 10 (a). Of course, the breakdown at the CCR-to-guard ring region cannot be avoided due to the CCR electrode n-type implant (See Fig. 10 (b).) Of course, the 10 $\mu m$ p-stop design was also shown to breakdown at much higher bias than measurement results. Thus, we can expect the breakdown would be pushed down to around $V_{bulk}$ = -400 $V$.

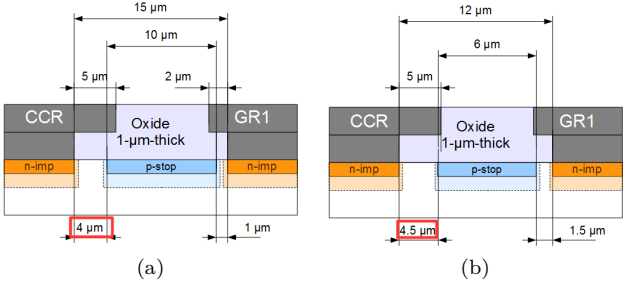

Fig. 9: (a) CCR-to-guard ring vicinity for current design (10 $\mu m$-p-stop-width.) (b) 6 $\mu m$-p-stop-width design.

In fact, we have already tested similar structure with 5 $\mu m$ spacing from n-type contact as depicted in Fig. 4, the "CCR Float" case. The distance between the outermost pixel electrode contact and the CCR contact is 15 $\mu m$ and the 5-$\mu m$-wide p-stop implant was placed at the middle of the Pixel and CCR spacing. In this case, we expect the CCR to be the innermost guard ring and the 5-$\mu m$-wide p-stop becomes the first guard ring implant. Thus, if the guard ring breakdown is solely dependent on the distance of the first guard ring p-type implant, $V_{bulk}$ = -240 $V$ will be the worst case scenario.

The surface breakdown at N-i-P (CCR implant to the first guard ring) was spotted when the potential drop at the N-i-P junction was around -75 $V$ which is higher than the 10 $\mu m$ case as shown in Fig. 11 (a). The 0.5 $\mu m$ of extra space between the n-type and p-type implants obviously improves the breakdown strength. However, the shrunken guard ring spacing limits the total potential drop up to around $V_{bulk}$ = -775 $V$ range while the guard ring breakdown was observed under $V_{bulk}$ = -1350 $V$. In other words, the depletion region reaches out to the wafer termination, rendering the guard ring breakdown out of scope from maximum operation bias of $V_{bulk}$ = -775 $V$. In other words, the simulation results are not trustworthy above the maximum operation bias, since the simulation model is missing the trap states at the wafer termination.

Indeed, the intensity of lateral electric field at the guard ring breakdown is higher than 10 $\mu m$ case due to higher negative $V_{bulk}$ bias than the original design, as depicted in Fig. 11 (b). It can also be noted that the actual electric field maximum is found at the 6th p-stop guard ring implant rather than the first guard ring implant, which is responsible for guard ring breakdown, yet the electric field strength diminishes after the 6th guard ring. The electric field intensity of the first guard ring at breakdown is around 0.34 $MV/cm$ which is, indeed, higher than the 10 $\mu m$ p-stop design. However, such high electric field at the 4th and 6th guard rings, which has the same distance from adjacent n-type implants due to the constant outwards overhang length, indicates that the distance from the previous contact needs to be adjusted accordingly on each guard ring.

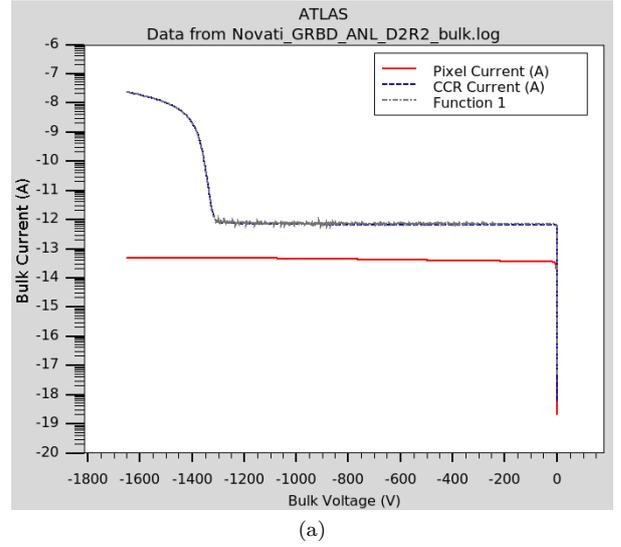

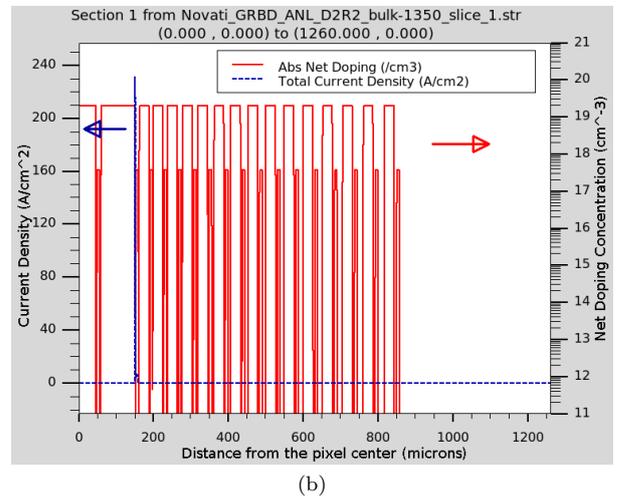

Fig. 10: (a) Improved breakdown characteristics from revised device and (b) the breakdown current profile at the guard ring junctions in the revised device. Again, the "Function 1" is the absolute value of bulk current: $|I_{bulk}|$.

## V. CONCLUSION

In this work, we investigated guard ring design parameters for N-in-P silicon sensors to eliminate plasma effects from the FASPAX high speed pixellated integrating detector for high intensity synchrotron X-ray light source. The TCAD simulation indicates that the first guard ring p-type implant must be at least 4.5 $\mu m$ away from n-type implant, with the given doping concentration profile, to avoid such unpredicted guard ring breakdown within the operational bias. In addition, we predict the maximum

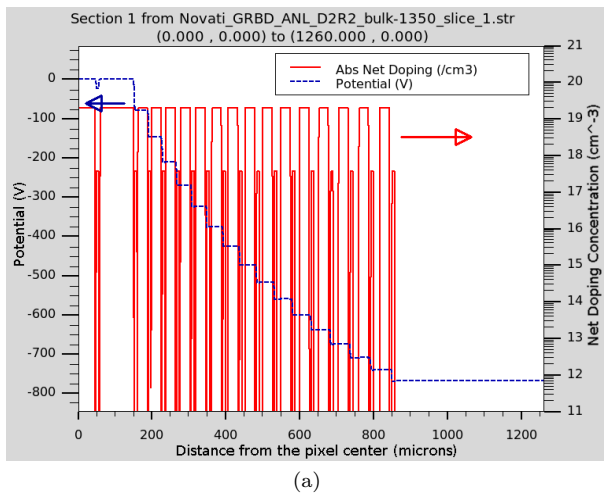

(a)

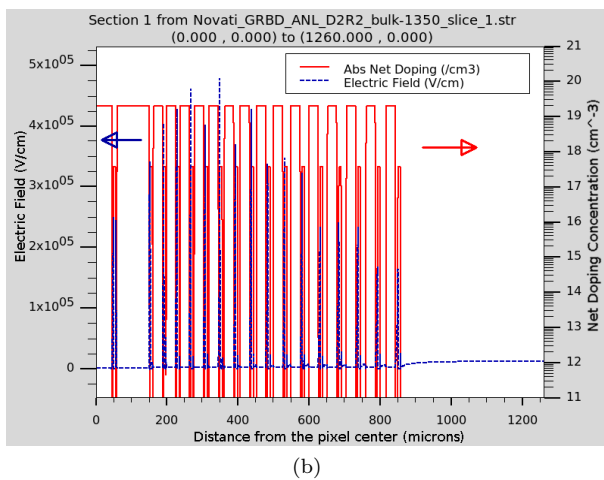

(b)

Fig. 11: (a) Surface potential drop of the revised device and (b) electric field at guard ring implants.

operational bias for the FASPAX silicon detector with revised design as -775 $V$.

The revised design has been taped out and sent to the Novati Technologies Inc. for a second batch of fabrication and is expected to be receeived in early 2016. These second sensors will be chracterized both electrically and with pulsed IR laser irradiation to determine the detector operational bias to prevent the plasma delay effect.

### Acknowledgment

Work at Argonne National Laboratory was supported by the U. S. Department of Energy, Office of Science, Office of Basic Energy Sciences, under Contract No. DE-AC02-06CH1135.

### References

[1] A. Thompson, F. Goulding, H. Sommer, J. Walton, E. Hughess, J. Rolfe, and H. Zeman, "A Multi-Element Silicon Detector for X-Ray Flux Measurements," *Nuclear Science, IEEE Transactions on*, vol. 29, no. 1, pp. 793–797, Feb. 1982.

[2] J. Kemmer, P. Burger, R. Henck, and E. Heijne, "Performance and Applications of Passivated Ion-Implanted Silicon Detectors," *Nuclear Science, IEEE Transactions on*, vol. 29, no. 1, pp. 733–737, Feb. 1982.

[3] E. Gatti and P. Rehak, "Semiconductor drift chamber An application of a novel charge transport scheme," *Nuclear Instruments and Methods in Physics Research*, vol. 225, no. 3, pp. 608 – 614, 1984.

[4] P. Fischer, J. Hausmann, M. Overdick, B. Raith, N. Wermes, L. Blanquart, V. Bonzom, and P. Delpierre, "A counting pixel readout chip for imaging applications," *Nuclear Instruments and Methods in Physics Research Section A: Accelerators, Spectrometers, Detectors and Associated Equipment*, vol. 405, no. 1, pp. 53 – 59, 1998.

[5] M. Campbell, E. Heijne, G. Meddeler, E. Pernigotti, and W. Snoeys, "A readout chip for a 64 times;64 pixel matrix with 15-bit single photon counting," *Nuclear Science, IEEE Transactions on*, vol. 45, no. 3, pp. 751–753, Jun. 1998.

[6] K. Taguchi and J. S. Iwanczyk, "Vision 2020: Single photon counting x-ray detectors in medical imaging," *Medical Physics*, vol. 40, no. 10, p. 100901, 2013.

[7] Argonne National Laboratory, "A MBA Lattice at the APS: A New Generation," 2013. [Online]. Available: http://aps.anl.gov/Upgrade/Workshops/2013/MBA-Technology/includes/aps%20workshop%20report.pdf

[8] K. Karim, "Active Matrix Flat Panel Imagers (AMFPI) for Diagnostic Medical Imaging Applications," in *Medical Imaging*. John Wiley & Sons, Inc., 2009, pp. 23–58.

[9] C.-y. Chen and J. Kanicki, "High field-effect-mobility a-Si:H TFT based on high deposition-rate PECVD materials," *Electron Device Letters, IEEE*, vol. 17, no. 9, pp. 437–439, Sep. 1996.

[10] A. Mozzanica, A. Bergamaschi, R. Dinapoli, H. Graafsma, B. Henrich, P. Kraft, I. Johnson, M. Lohmann, B. Schmitt, and X. Shi, "A single photon resolution integrating chip for microstrip detectors," *Nuclear Instruments and Methods in Physics Research Section A: Accelerators, Spectrometers, Detectors and Associated Equipment*, vol. 633, Supplement 1, pp. S29 – S32, 2011, 11th International Workshop on Radiation Imaging Detectors (IWORID).

[11] A. A. Quaranta, A. Taroni, and G. Zanarini, "Plasma time and related delay effects in solid state detectors," *Nuclear Instruments and Methods*, vol. 72, no. 1, pp. 72 – 76, 1969.

[12] G. Ottaviani, C. Canali, and A. Quaranta, "Charge Carrier Transport Properties of Semiconductor Materials Suitable for Nuclear Radiation Detectors," *Nuclear Science, IEEE Transactions on*, vol. 22, no. 1, pp. 192–204, Feb. 1975.

[13] Z. Sosin, "Description of the plasma delay effect in silicon detectors," *Nuclear Instruments and Methods*

*in Physics Research Section A: Accelerators, Spectrometers, Detectors and Associated Equipment*, vol. 693, pp. 170 – 178, 2012.

[14] J. Becker, K. Gärtner, R. Klanner, and R. Richter, "Simulation and experimental study of plasma effects in planar silicon sensors," *Nuclear Instruments and Methods in Physics Research Section A: Accelerators, Spectrometers, Detectors and Associated Equipment*, vol. 624, no. 3, pp. 716 – 727, 2010.

[15] P. Hart, S. Boutet, G. CarmI, A. Dragone, B. Duda, D. Freytag, G. Haller, R. Herbst, S. Herrmann, C. Kenney, J. Morse, M. Nordby, J. Pines, N. van Bakel, M. Weaver, and G. Williams, "The Cornell-SLAC pixel array detector at LCLS," in *Nuclear Science Symposium and Medical Imaging Conference (NSS/MIC), 2012 IEEE*, Oct. 2012, pp. 538–541.

[16] Y. Kao and E. Wolley, "High-voltage planar p-n junctions," *Proceedings of the IEEE*, vol. 55, no. 8, pp. 1409–1414, Aug. 1967.

[17] L. Evensen, A. Hanneborg, B. S. Avset, and M. Nese, "Guard ring design for high voltage operation of silicon detectors," *Nuclear Instruments and Methods in Physics Research Section A: Accelerators, Spectrometers, Detectors and Associated Equipment*, vol. 337, no. 1, pp. 44 – 52, 1993.

[18] N. Bacchetta, M. Da Rold, R. Dell'Orso, P. Fuochi, A. Lanza, A. Messineo, O. Militaru, A. Paccagnella, G. Verzellesi, and R. Wheadon, "Study of breakdown effects in silicon multi-guard structures," in *Nuclear Science Symposium, 1997. IEEE*, Nov. 1997, pp. 498–502 vol.1.

[19] D. C. Sheridan, G. Niu, J. N. Merrett, J. D. Cressler, C. Ellis, and C.-C. Tin, "Design and fabrication of planar guard ring termination for high-voltage SiC diodes," *Solid-State Electronics*, vol. 44, no. 8, pp. 1367 – 1372, 2000.

[20] M. Lozano, G. Pellegrini, C. Fleta, C. Loderer, J. Rafi, M. Ullan, F. Campabadal, C. Martinez, M. Key, G. Casse, and P. Allport, "Comparison of radiation hardness of P-in-N, N-in-N, and N-in-P silicon pad detectors," *Nuclear Science, IEEE Transactions on*, vol. 52, no. 5, pp. 1468–1473, Oct. 2005.

[21] Y. Unno, A. A. Affolder, P. P. Allport, R. Bates, C. Betancourt, J. Bohm, H. Brown, C. Buttar, J. R. Carter, G. Casse, H. Chen, A. Chilingarov, V. Cindro, A. Clark, N. Dawson, B. DeWilde, Z. Dolezal, L. Eklund, V. Fadeyev, D. Ferrere, H. Fox, R. French, C. Garcia, M. Gerling, S. G. Sevilla, I. Gorelov, A. Greenall, A. A. Grillo, N. Hamasaki, K. Hara, H. Hatano, M. Hoeferkamp, L. B. A. Hommels, Y. Ikegami, K. Jakobs, S. Kamada, J. Kierstead, P. Kodys, M. Kohler, T. Kohriki, G. Kramberger, C. Lacasta, Z. Li, S. Lindgren, D. Lynn, M. Mikestikova, P. Maddock, I. Mandic, S. M. i. Garcia, F. Martinez-McKinney, R. Maunu, R. McCarthy, J. Metcalfe, M. Mikuz, M. Minano, S. Mitsui, V. O'Shea, S. Paganis, U. Parzefall, D. Puldon, D. Robinson, H. F.-W. Sadrozinski, S. Sattari, D. Schamberger, S. Seidel, A. Seiden, S. Terada, K. Toms, D. Tsionou, J. V. Wilpert, M. Wormald, J. Wright, M. Yamada, and K. Yamamura, "Development of n-on-p silicon sensors for very high radiation environments," *Nuclear Instruments and Methods in Physics Research Section A: Accelerators, Spectrometers, Detectors and Associated Equipment*, vol. 636, no. 1, Supplement, pp. S24 – S30, 2011, 7th International.

[22] C. Gallrapp, A. L. Rosa, A. Macchiolo, R. Nisius, H. Pernegger, R. H. Richter, and P. Weigell, "Performance of novel silicon n-in-p planar pixel sensors," *Nuclear Instruments and Methods in Physics Research Section A: Accelerators, Spectrometers, Detectors and Associated Equipment*, vol. 679, pp. 29 – 35, 2012.

[23] Silvaco Inc., "Device 3d: 3d Device Simulator Brochure," 2015. [Online]. Available: http://www.silvaco.com/products/tcad/device_simulation/atlas/atlas.html

[24] J. Becker, L. Bianco, P. Gottlicher, H. Graafsma, H. Hirsemann, S. Jack, A. Klyuev, A. Marras, U. Trunk, R. Klanner, J. Schwandt, J. Zhang, R. Dinapoli, D. Greiffenberg, B. Henrich, A. Mozzanica, B. Schmitt, X. Shi, M. Gronewald, and H. Kruger, "Architecture and design of the AGIPD detector for the European XFEL," in *Nuclear Science Symposium and Medical Imaging Conference (NSS/MIC), 2012 IEEE*, Oct. 2012, pp. 585–590.

[25] Silvaco Inc., "DevEdit User's Manual," Jan. 2014. [Online]. Available: https://dynamic.silvaco.com/dynamicweb/jsp/downloads/DownloadManualsAction.do?req=silen-manuals&nm=devedit

[26] ——, "Atlas User's Manual," Oct. 2015. [Online]. Available: https://dynamic.silvaco.com/dynamicweb/jsp/downloads/DownloadManualsAction.do?req=silentmanuals&nm=atlas

[27] J. G. Fossum and D. S. Lee, "A physical model for the dependence of carrier lifetime on doping density in nondegenerate silicon," *Solid-State Electronics*, vol. 25, no. 8, pp. 741 – 747, 1982.

[28] P. Nayak, "Characterization of High-Resistivity Silicon Bulk and Silicon-on-Insulator Wafers," Ph.D. dissertation, Arizona State University, Aug. 2011. [Online]. Available: http://hdl.handle.net/2286/R.I.15784

[29] Aurientis, Martin, "Interdigitated Back Contact n-Type Solar Cell with Black Silicon Anti-Reflecting Layer: Simulations and Experiments," Ph.D. dissertation, Aalto University, Aug. 2014. [Online]. Available: http://urn.fi/URN:NBN:fi:aalto-201409012586